\documentclass[]{spie}  

\usepackage{amsmath,amsfonts,amssymb}
\usepackage{graphicx}
\usepackage{subfigure}
\usepackage{comment}
\usepackage{marginnote}
\usepackage[colorlinks=true, allcolors=blue]{hyperref}
\usepackage[table,xcdraw]{xcolor}
\usepackage[normalem]{ulem}

\title{Pyxel: the collaborative detection simulation framework}

\author[a]{Thibaut Prod'homme}
\author[a]{Fr\'ed\'eric Lemmel}
\author[a]{Matej Arko}
\author[b]{Benoit Serra}
\author[b]{Elizabeth George}
\author[c]{Enrico Biancalani}
\author[a]{Hans Smit}
\author[d]{David Lucsanyi}

\affil[a]{European Space Agency, ESTEC, Keplerlaan 1, 2201 AZ, Noordwijk, The Netherlands}
\affil[b]{European Southern Observatory, Karl-Schwarzschild-Str. 2, 85748 Garching bei München, Germany}
\affil[c]{Leiden Observatory, Leiden University, Niels Bohrweg 2, 2333 CA Leiden, The Netherlands}
\affil[d]{CERN, Esplanade des Particules 1, 1217 Meyrin, Switzerland}

\authorinfo{Send Pyxel user request and Pyxel-related questions to: pyxel@esa.int. For contacting the authors send correspondence to: thibaut.prodhomme@esa.int \\ Paper to be published as part of the SPIE Astronomical Telescopes and Instrumentation 2020 conference proceedings.}

\begin{document}

\maketitle

\begin{abstract}
Pyxel is a novel python tool for end-to-end detection chain simulation i.e. from detector optical effects to readout electronics effects. It is an easy-to-use framework to host and pipeline any detector effect model. It is suited for simulating both Charge-Coupled Devices, CMOS Image Sensors and Mercury Cadmium Telluride hybridized arrays. It is conceived as a collaborative tool to promote reusability, knowledge transfer, and reliability in the instrumentation community. We will provide a demonstration of Pyxel’s basic principles, describe newly added capabilities and the main models already implemented, and give examples of more advanced applications. 
\end{abstract}


\section{INTRODUCTION}
\label{sec:intro} 

\begin{figure} [!ht]
   \begin{center}
   \begin{tabular}{c}
     \includegraphics[height=3cm]{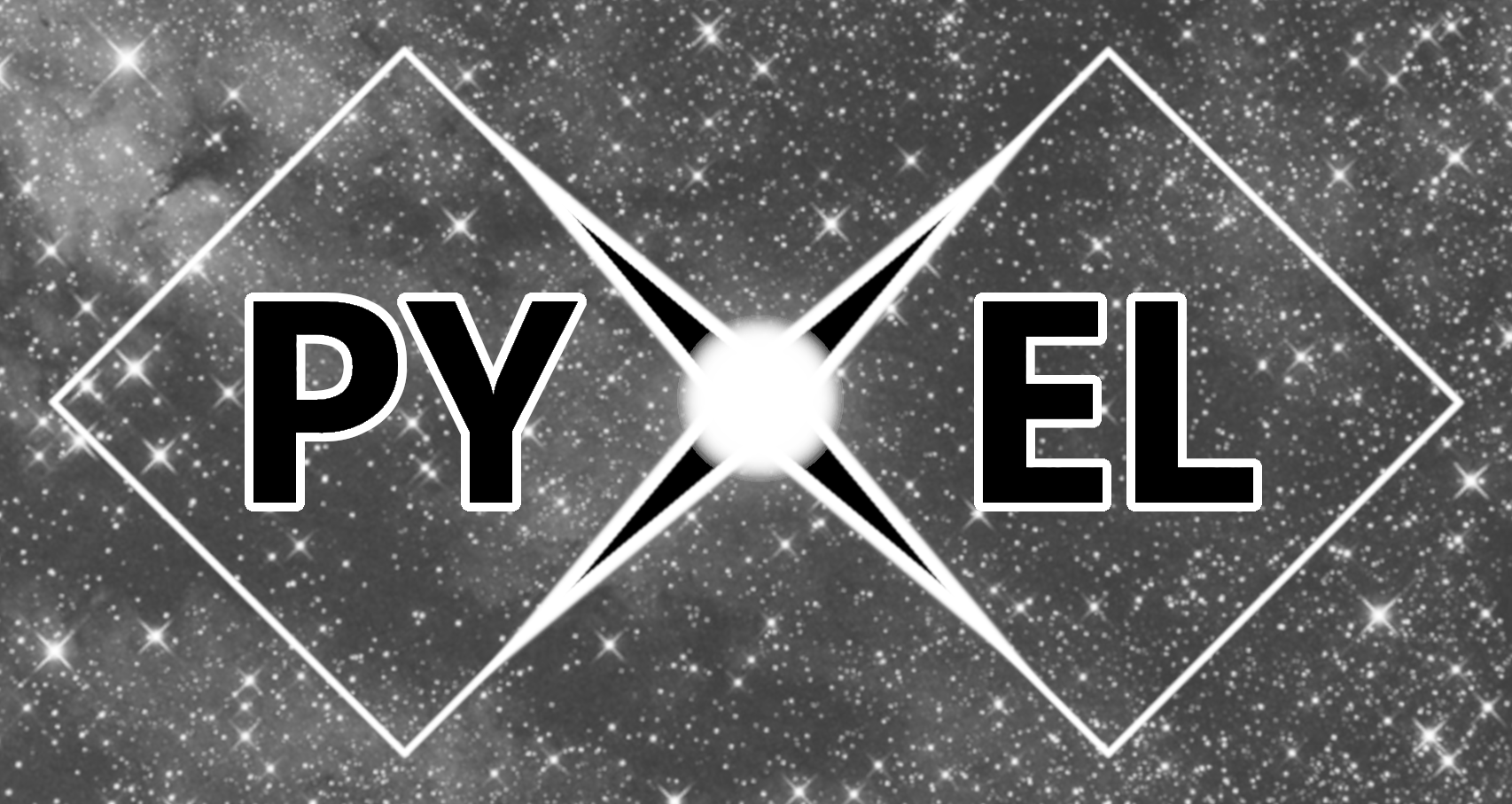}
   \end{tabular}
   \end{center}
   \caption[logo] 
   { \label{fig:logo}  \centering
	The logo of Pyxel, the open-source detector simulation framework.}
   \end{figure} 

Due to their size, complexity, and very stringent scientific requirements, the current and next generation of instruments on-board ESA Science and Earth-observation missions and the ones developed for the ground-based Extremely Large Telescopes in particular at ESO are pushing technical capabilities ever closer to their theoretical limits. In parallel, rapid advances in computational methods have unlocked the capability of relying on instrument simulations accounting for complex optical, mechanical, thermal, and other effects to achieve these high-precision science cases. Modelling has therefore become a key element in the development and operation of scientific instruments. It now truly spans all phases of a project: translating high-level science requirements into a defined instrument concept, supporting technology selection and trade-off analyses during the definition and design phases, monitoring compliance between science performance specifications and requirements during the  design and construction phases, optimizing science return through calibration, planning for science operations, and processing and analyzing the final data products.

One critical component of instrument simulation is the behavior of the detectors. While many space- and ground-based projects use similar (and in some cases identical) detectors, most projects have implemented basic detector effects into their instrument simulators independently. Furthermore despite the wealth of detector effect models described in the literature, the models' source code is not always available, their implementation can be non-trivial, and the models cannot be easily combined to form a complete detector simulation. In a joint-effort to circumvent these limitations, ESA and ESO have developed a collaborative detection simulation framework called Pyxel. Pyxel is a python tool for end-to-end detection chain simulation i.e. from detector optical effects to readout electronics effects. It is an easy-to-use framework to host and pipeline any detector effect model. It is suited for simulating a variety of detectors; it has been used so far to simulate traditional sensors such as CCDs (Charge-Coupled Devices), CMOS Image Sensors, Mercury Cadmium Telluride hybridized arrays (referred to as MCT in the following), and more recently photon-counting MKID-arrays. Pyxel is open-source and conceived as a tool to enable and foster collaborations in between organizations and individuals in the instrumentation community. It promotes reusability, knowledge transfer, and reliability. 

Pyxel was first presented\cite{pyxel1} to the community two years ago and has raised a lot of interest. In the mean time it has been developed further to reach a stable state with a focus on establishing a robust but flexible and easy-to-use framework and improve the user experience. 
A first beta release took place in early 2019. Since then, a growing community of early users has added new models and helped the main developers at ESA and ESO to develop new features e.g. the Jupyter notebook interface and to polish existing features for greater usability such as: the parametric running mode to allow model parameter sensitivity analysis, the calibration running mode to validate and calibrate models against experimental test data, and the dynamic running mode to enable the simulation of time-dependent operations like non-destructive readout mode for MCT devices or Time-delayed Integration for CCDs. Last but not least, recent efforts have enabled the use of multi-threaded and parallel computing; Pyxel is now able to run on computer clusters and the cloud to perform heavy simulations, generate large synthetic datasets, and solve complex optimization problems.

In this contribution we first give an updated summary of Pyxel's main working principles from its architecture to the current list of models included. We then provide a short guide on how to install and use Pyxel followed by examples of simulations one can perform. Ultimately we conclude with an outlook towards v1.0.

\section{Pyxel in a nutshell}
\label{sec:summary}

The main programming language used for Pyxel is Python 3. This programming language was selected for its cross-platform capability and its popularity; Python has established itself as the ``lingua franca'' among scientists and software developers in physical and data science. As a result, Pyxel is able to run seamlessly on Linux-based, Windows and macOS operating systems.

Following Pyxel's founding principles of promoting reusability and reliability, the software is based as much as possible on well-known modern Python open source libraries used in the scientific community such as NumPy \cite{numpy}, SciPy \cite{scipy}, Pandas \cite{pandas}, Matplotlib \cite{matplotlib}, Astropy \cite{astropy}, Jupyter Notebooks \cite{jupyter}, Numba \cite{numba} and Dask \cite{dask}.

Pyxel is also open-source and readily available - for now on user request - from its gitlab repository \cite{pyxelgitlab}. Pyxel licensing process is on-going, the chosen license is the well-known permissive MIT license (expat) enabling unrestricted commercial use, distribution, modification, and private use by tiers. Once its licensing process is achieved, Pyxel shall be available directly from commonly-used software repositories such as PyPI (the Python Package Index). 

\subsection{Architecture}
\begin{figure} [ht]
   \begin{center}
   \begin{tabular}{c}
     \includegraphics[width=0.7\columnwidth]{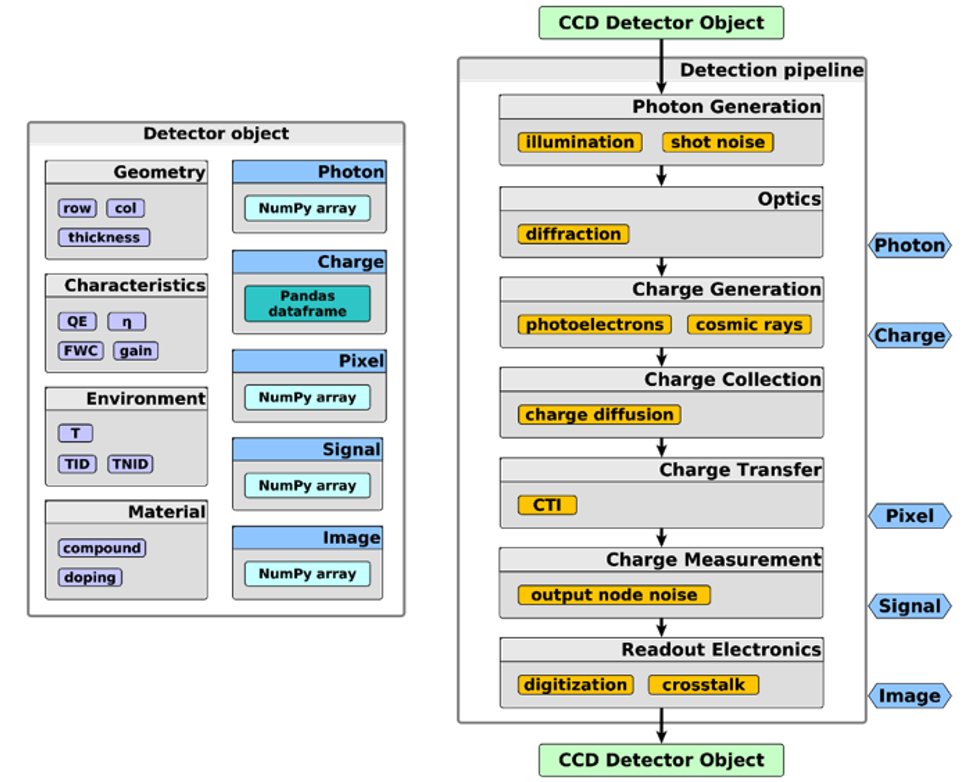}
   \end{tabular}
   \end{center}
   \caption[architecture] 
   { \label{fig:architecture}  \centering
	Graphical representation of Pyxel's generic detector object (left) together with an example of a detection pipeline for CCD simulation (right).}
\end{figure} 

The overall software architecture of Pyxel is best described by its three main elements:
\begin{itemize}
    \item the detection pipeline: the core algorithm, hosting and running the detector effect models, which are grouped into different levels imitating the working principles of the a detector (see Fig.~\ref{fig:architecture} right).
    \item the detector object: a bucket containing all properties and data of the simulated detector, which is passed to all the models activated by the user, and may be used or modified by them as well (see Fig.~\ref{fig:architecture} left).
    \item the configuration file: Pyxel's user unique entry point. Based on YAML\cite{yaml}, the configuration file is structured, easy-to-read and understand. Fig.~\ref{fig:yaml_configuration} shows an example, and section~\ref{sec:usecases} gives more details on how it is structured.
    
\end{itemize}
A more complete and in-depth description of Pyxel's architecture can be found here\cite{pyxel1}.

\subsection{Models}
\label{sec:models}

Models are Python functions with a Detector object defined as their input argument. The model function has to be added to the YAML configuration file. Then the function is automatically called by Pyxel inside a loop of its model group and the Detector object is passed to it. The model may modify this Detector object which is then subsequently used and modified further by the next models in the pipeline.

Currently implemented models vary from very simple analytical functions, e.g. the simple Rule07 equation\cite{rule07} for dark current theoretical predictions in MCT devices, to more complex algorithms such as CosmiX\cite{cosmix} a model to simulate charge deposition by cosmic rays in detectors or the recently published Tulloch and George\cite{tulloch} model for persistence in MCT devices (see an example of use of this model in Section~\ref{sec:usecases}).

Pyxel's website\cite{pyxelwebsite} lists a comprehensive list of models currently implemented and under implementation with references, it also provides a concise how-to on adding models to the framework.

\subsection{Running modes}
\label{sec:running modes}
Pyxel can be run in four different modes:
\begin{itemize}
   \item Single: ``one image in, one image out'', a single pipeline run to have quick look and make a health check.
   \item Parametric: the detection pipeline is run multiple times looping over a range of model or detector parameters. This mode is particularly useful to run sensitivity analysis or to generate large image datasets.
   \item Calibration: using a user-defined fitness function and built-in optimization algorithms (e.g. genetic algorithms\cite{pygmo}) this mode optimizes model or detector parameters to fit a user-defined target dataset. This mode can be used to find the best instrument operating parameters or to calibrate models against experimental data. 
   \item Dynamic: the pipeline is run $N$ times in a loop and the output detector becomes the input detector for each iteration and increment in time. This mode is key to simulate non-destructive readout mode in MCT devices and time-dependent effects such as persistence.
\end{itemize}

\subsection{Collaboration}
One focus of Pyxel since its beta release in early 2019 has been to welcome not only new users, but also new developers. To enable and support a growing Pyxel collaboration, on top of the already existing gitlab repository\cite{pyxelgitlab}, we have developed the following tools: (i) a website\cite{pyxelwebsite} hosting both Pyxel's improved and comprehensive documentation and a blog to share the latest news, developments, and advertise open positions and the improved documentation, (ii) a mailing list\cite{mailinglist} and a chat platform\cite{chat} where users can reach out to the community for technical support and sharing experience, tips, hacks, (iii) an improved installation guide to simplify the installation process for Pyxel newcomers, (iv) a detailed contribution guide to help more advanced users contributing to Pyxel via Gitlab e.g. improving documentation, adding new models, reporting bugs.

\section{Using Pyxel}
\label{sec:usecases}

In this section we first briefly explain how to install Pyxel and how to use it. We then give two examples of recent use cases of Pyxel; namely, simulating photon-counting MKID-arrays, and simulating persistence in HxRG-like detectors.

\subsection{How to install Pyxel?}

The installation of Pyxel was simplified for developers and contributors. A detailed guide available on the website explains: (i) how to fork the project Pyxel on Gitlab\cite{gitlab}, (ii) how to get the source code on one's computer and (iii) how to prepare a development environment on Python and Anaconda.

For the end users that only want to use Pyxel, it will soon be possible to easily deploy this tool through the PyPI or Anaconda repositories. Pyxel will be available on Python 3.7+ for Windows, MacOS and Linux. A docker image including a Jupyter Notebook server will be also available soon.

It will be also possible to run Pyxel from everywhere and without installation directly from Binder \cite{binder}. This mode of operation will be mainly used to run examples and tutorials.

\subsection{How to use Pyxel?}

The main user input is a user-friendly YAML configuration file, which defines all the necessary parameters to run a simulation. It consists of three sections: (i) running mode definition, (ii) detector properties definition, (iii) detection pipeline construction. The structure of the configuration file  mimics the structure and hierarchy of classes that are initialized from it at the very start of the program. Loading the configuration file instantiates the classes describing the detector with unique properties (environment, geometry, characteristics and material), the detection pipeline class hosting model groups and model functions, and the class describing the running mode. The newly created running mode object stores the user defined outputs as well as all additional arguments that may be needed for different modes. These can be for example parameters for parametric mode, fitting function and algorithm for calibration mode, time scale for dynamics, etc. One such configuration file for parametric mode is shown on Fig.~\ref{fig:yaml_configuration}.

On top of running the Pyxel scripts directly from an IDE, users can also run Pyxel either as a standalone application locally from the command line, in the so-called batch mode, or as a library and import it elsewhere in a Jupyter notebook for instance. In the command line, only a filename of the configuration must be passed as an argument:

\begin{verbatim}
$ pyxel --config input.yaml
\end{verbatim}

\begin{figure} [!ht]
   \begin{center}
   \includegraphics[width=0.99\columnwidth]{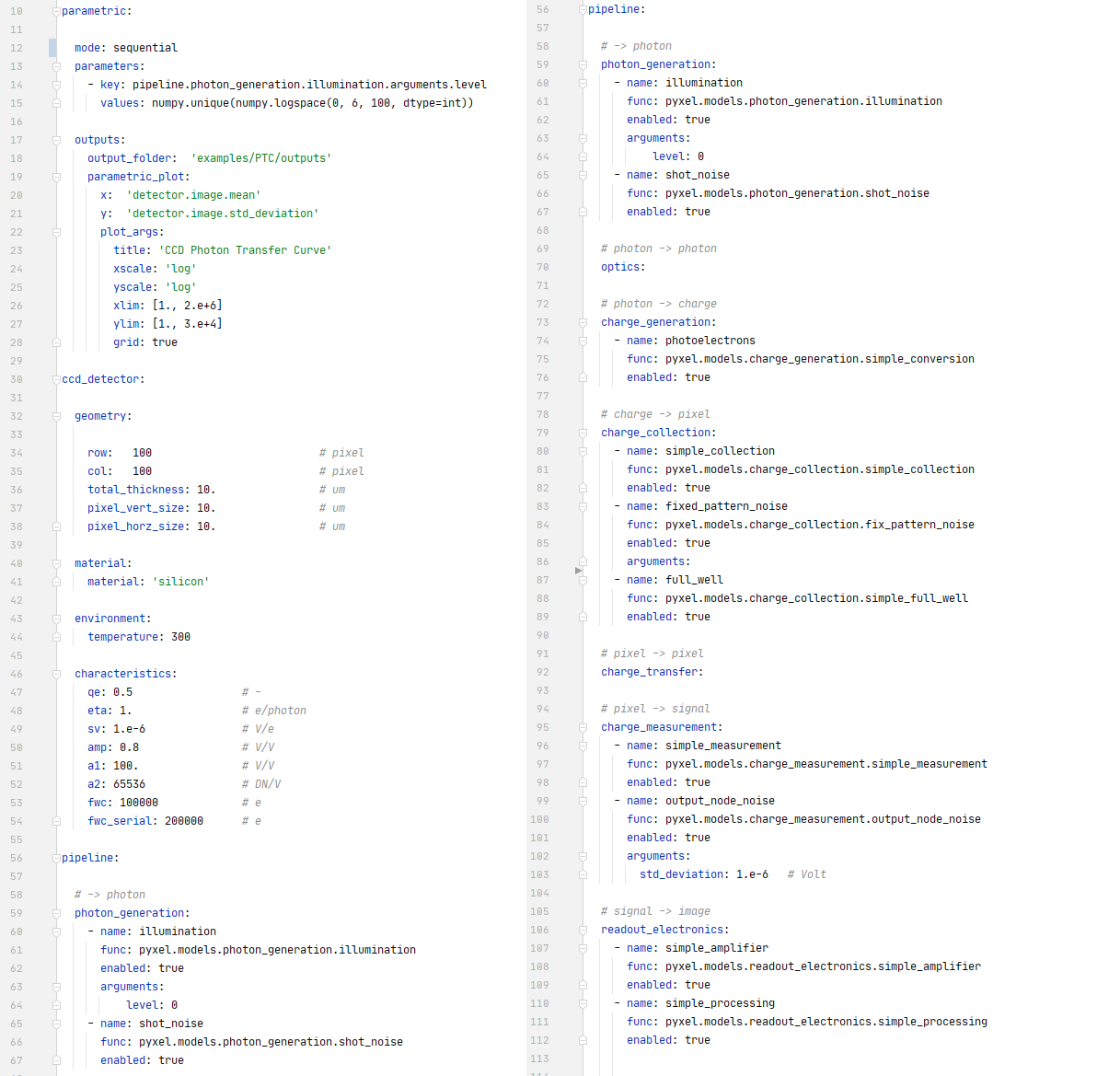}
   \caption{Configuration file for the parametric mode simulation in one of the Pyxel usage examples\cite{pyxeldata}. The configuration consists of three sections, each one representing an object in the Pyxel architecture. The running mode class, in this case the mode is parametric, also holds information about the outputs, the detector class stores the detector properties and the detection pipeline class stores all the model groups with the model functions.}
   \label{fig:yaml_configuration}
   \end{center}
\end{figure} 

\begin{figure} [ht]
   \begin{center}
   \includegraphics[width=0.9\columnwidth]{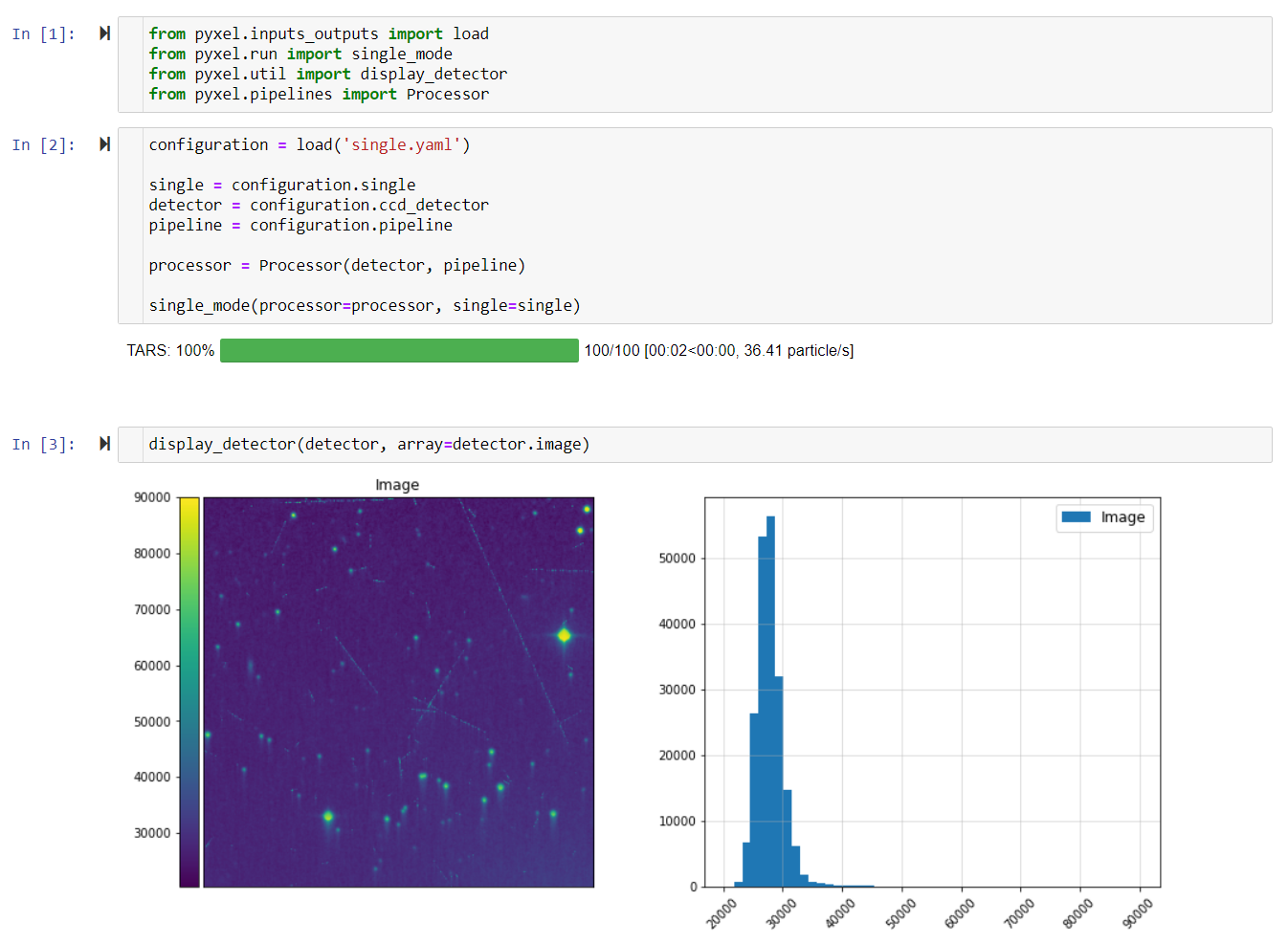}
   \caption{The simplest example of running the single mode simulation in the Jupyter notebook. After the creation of classes from the three sections of the input YAML file, processor object can be initialized. By running the single mode simulation, the detector object passes through the pipeline and the result of the simulated image can be displayed at the end. Here a CCD acquisition is simulated with in particular visible the effects of the radiation-induced charge transfer inefficiency (trails opposite to the transfer direction) and cosmic ray charge deposition (long streaks).}
   \label{fig:jupyter}
   \end{center}
\end{figure} 

Additional optional arguments are available, such as verbosity or random seed.
The Jupyter Notebook environment is easy to use, has a large user base and offers a wide variety of visualisation and interactivity possibilities. It provides a more robust, reliable, and maintainable user interface than an ad-hoc developed GUI (Graphical User Interface). For this reason we halted the development of the Pyxel GUI and focus now on improving the integration of Pyxel to Jupyter notebooks for instance by the addition of visualisation options; Pyxel usage examples are now made available as Jupyter notebooks. One such example is shown in Fig.~\ref{fig:jupyter}. The notebook examples are stored in a separate public Gitlab repository\cite{pyxeldata}, where one can find at least one working notebook per operating mode, together with the example YAML files.

Furthermore, any examples or tutorials provided in Pyxel can also be executed from anywhere and without installation thanks to Binder.
Binder will automatically create a new Jupyter Notebook environment with Pyxel installed and will allow to run immediately any pipelines. 

For CPU intensive operations in parametric and calibration modes, there is a possibility of executing both modes in parallel on a single computer, on a grid of computers working as a job queuing system like SGE\cite{SGE}, or in a cloud of computers. 
The parametric mode can be executed in parallel with Dask and the calibration mode can be executed on grid of computers with ipyparallel\cite{ipyparallel}.
Later, Dask will be used for any parallel computing.

\subsection{Model Examples}
There are currently many models implemented into Pyxel. Here we describe two of the latest additions, for two very different types of detectors. The first is a model of pixel dead-time in MKID arrays, and the second is a model of persistence in MCT detectors. In both cases, the models allow users to determine how these effects may impact their data, and therefore the observability of their science cases. 

\subsubsection{Simulating photon-counting MKID-arrays}

\begin{figure}[ht!]
\begin{center}

        \subfigure[]{
            \includegraphics[width=0.45\textwidth]{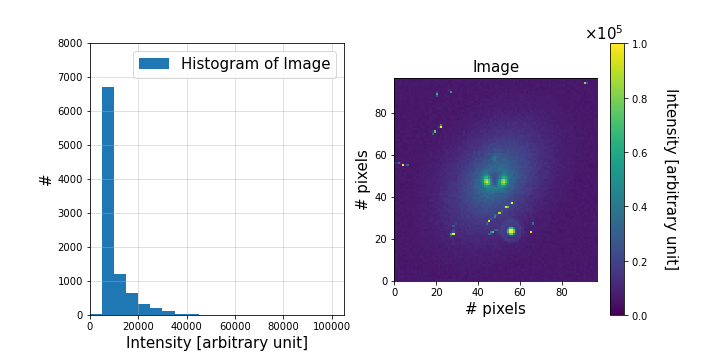}} 
        \subfigure[]{
            \includegraphics[width=0.45\textwidth]{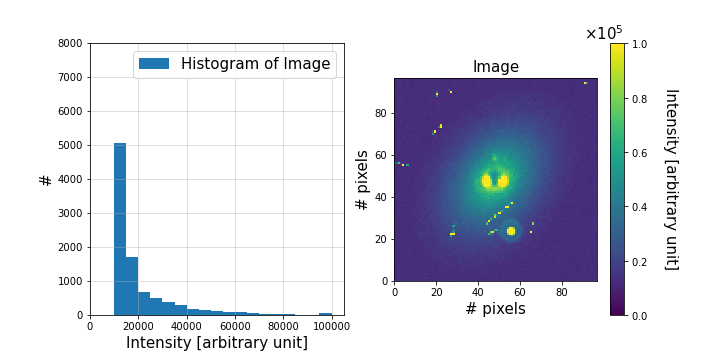}} \\
        \subfigure[]{
            \includegraphics[width=0.45\textwidth]{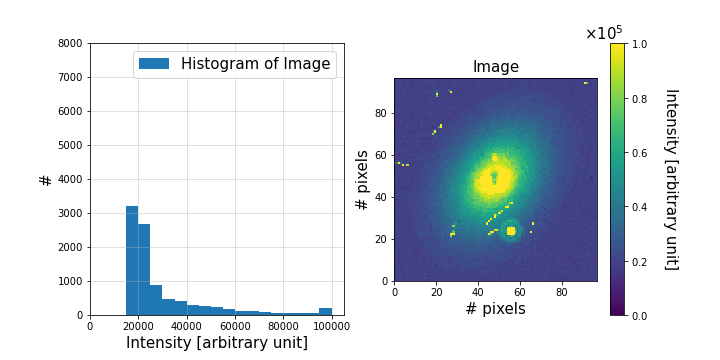}} 
        \subfigure[]{
            \includegraphics[width=0.45\textwidth]{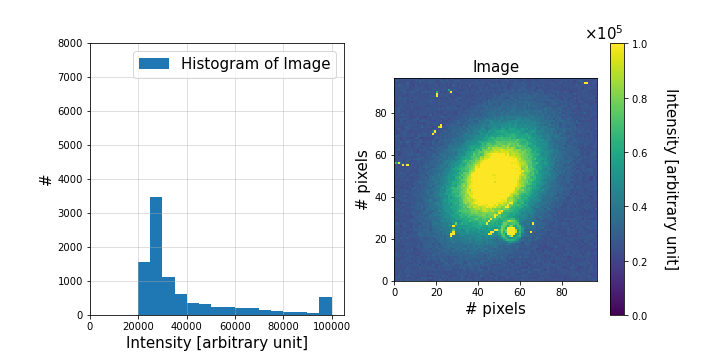}} \\
        \subfigure[]{
            \includegraphics[width=0.45\textwidth]{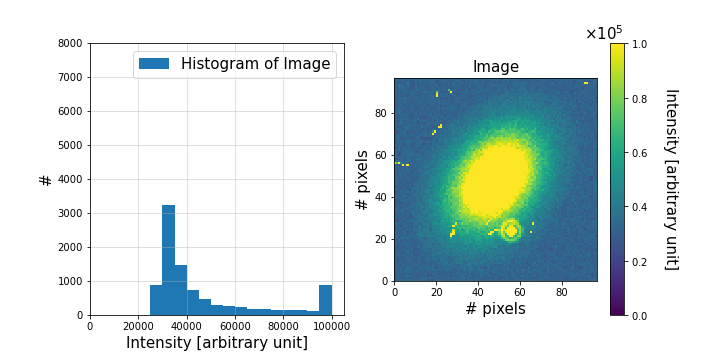}}
        \subfigure[]{
            \includegraphics[width=0.45\textwidth]{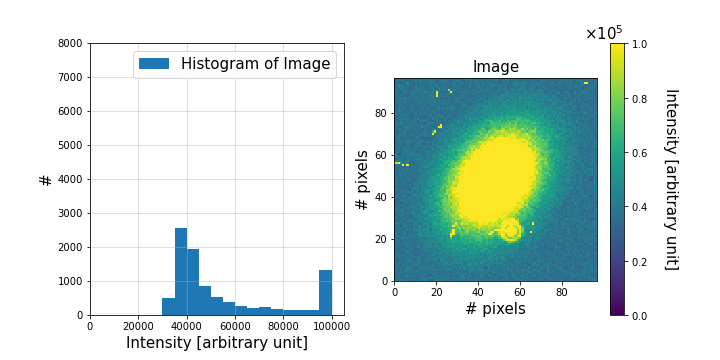}} 
            
        \caption{Mosaic of simulations showing the effect of temporal saturation for an MKID-array, which leads to an intensity saturation; by incrementally increasing the brightness level in the field of view, from (a) to (f). The effect appears when the interval between the arrival time of two photons is smaller than the dead time of the affected MKIDs in the array, assuming an ideal read-out bandwidth. The sequence of associated histograms shows how the counts ($\#$) move towards higher intensities, until the wall of 10$^{5}$ (in arbitrary units) is reached.}
        \label{fig:Enrico}

\end{center} 
\end{figure}

A superconducting microwave kinetic-inductance detector (MKID) is a novel concept of photo-detector tailored for wavelengths below a few millimetres \cite{Zmuidzinas}. Essentially, it is an LC micro-resonator with a superconducting photo-sensitive area, which is capable of detecting single photons; as well as providing a measurement on their arrival time and energy. Operationally, each photon impinging on the MKID breaks a number of superconducting charge carriers, called Cooper pairs, into quasi-particles. In turn, thanks to the principle of kinetic inductance---which becomes relevant only below a superconducting critical temperature---such a strike is converted into a sequence of microwave pulses (at a few $GHz$) with a wavelength-dependent profile; which are then read out one by one, on a single channel. An important caveat is that the photo-generated pulse profiles can be distinguished only if they do not overlap in time and if the read-out bandwidth is large enough. The situation is analogous to the ``pulse pile-up'' and ``coincidence losses'' of EM-CCDs, in photon-counting mode \cite{Wilkins}. In other words, there is a maximum limit to the achievable count rate, which is inversely proportional to the minimum distance in time between distinguishable pulse profiles: the so-called ``dead time'', which is fundamentally determined by the recombination time of quasi-particles re-forming Cooper pairs. Given this, an MKID-array per se can serve as an intrinsic integral-field spectrograph, at low resolution, without any dispersive elements or chromatic filters \cite{Mazin}.

In particular, at visible and NIR wavelengths, an MKID-based camera has the potential of revolutionising the low-resolution direct spectrography of exo-planets, thanks to the aforementioned characteristics of MKIDs, as well as their built-in high multiplexability \cite{Rauscher}; as long as the field of view is not too bright. The example in Fig. \ref{fig:Enrico} shows how Pyxel was used to simulate and investigate the effect of overlapping pulse profiles for an ideal MKID-array; assuming an ideal read-out bandwidth for such a science case. With the incrementally increasing brightness level of the field of view, this temporal saturation---for photons' arrival-time intervals smaller than the dead time---ends up affecting the contrast among the observables, thus leading to an intensity saturation. 

To test this saturation phenomenon, one can use Pyxel's dynamical simulation mode, by varying the parameter relative to the input illumination. Here, this corresponds to an exo-planetary system---pseudo-Fomalhaut star with injected Earth (top), resolvable only in (a), (b) and (c), and Jupiter (bottom)---observed by HabEx with a starshade, in a simulation performed with SISTER \cite{Hildebrandt}; plus ten cosmic rays added within Pyxel, which demonstrate the ease of pipelining different models even for a brand-new type of detector. Further characteristics of photon-counting MKID-arrays to be merged into Pyxel at a later stage can be found in the MEDIS simulator \cite{Dodkins}.

\subsubsection{Simulating persistence in HxRG-like detectors}

Persistence is one of the hottest topics at the moment for both space missions and ground-based experiments with sensing capabilities in the near-to-longwave infrared wavelength range, and for the detector manufacturers for such experiments. It is very important to characterize persistence, as it can in some extreme cases prevent observations for a long time after the detector is exposed to a bright source. Recently Tulloch and George\cite{tulloch} have developed a new model capable of explaining and reproducing the persistence effect in HxRG-like MCT detectors. This model was then implemented into Pyxel to probe this effect that can be seen both in the current exposure (trapping mechanism) and the subsequent exposures (detrapping).

To simulate persistence in Pyxel, we use the dynamic mode where the same detector object is used through several subsequent detection pipelines such that we can monitor the trapped charge throughout the simulation. For this, the detector object was modified to have a memory attribute (dictionary) which can be used to store temporary result such as the number of trapped charges.

First we simulate the illumination: the detector observed source is set to a bright object (the Pyxel logo here). The detector integrates signal for a number of steps until the exposure is complete. After the charge has been regrouped into one pixel: the charge number, the time since beginning of exposure, and the trap density are used to compute the number of trapped and detrapped charges. This number for each pixel is stored into the memory for each time constant (1, 10, 100, 1000, 10000) as a 2D map. Once the simulation is over, the detector object memory attribute contains the total number of trapped charges. 

\begin{figure}[ht!]
    \begin{center}
    \includegraphics[width=0.7\textwidth]{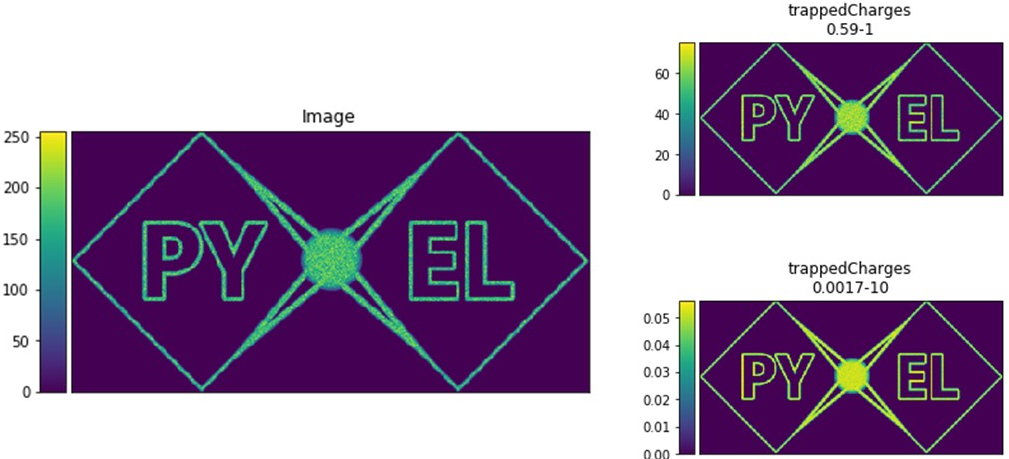}
    \caption{The detector object Image attribute (left) and memory (right) at the end of the exposure. The source is a FITS file of the Pyxel logo. The right top image is the trapped charge for the 1~s time constant, the right bottom one for the 10~s time constant.}
    \label{fig:persistence1}
\end{center} 
\end{figure}

The result of this first simulation is shown in Fig.~\ref{fig:persistence1}. For the memory of the detector, only the trapped charge maps for the first two time constants are displayed. As we increased the trap densities for the 1 second time constant, the first time constant traps dominate. Note that the chosen simulation parameters used here are not realistic, they have been tweaked in order to illustrate well the persistence in the figures. Trap densities for short time constants was multiplied by 100 compared to real detector measurements, and the exposure time was set to 10 seconds with one frame every 0.5 seconds. And all other models are disabled. 

\begin{figure}[ht!]
    \begin{center}
    \includegraphics[width=0.7\textwidth]{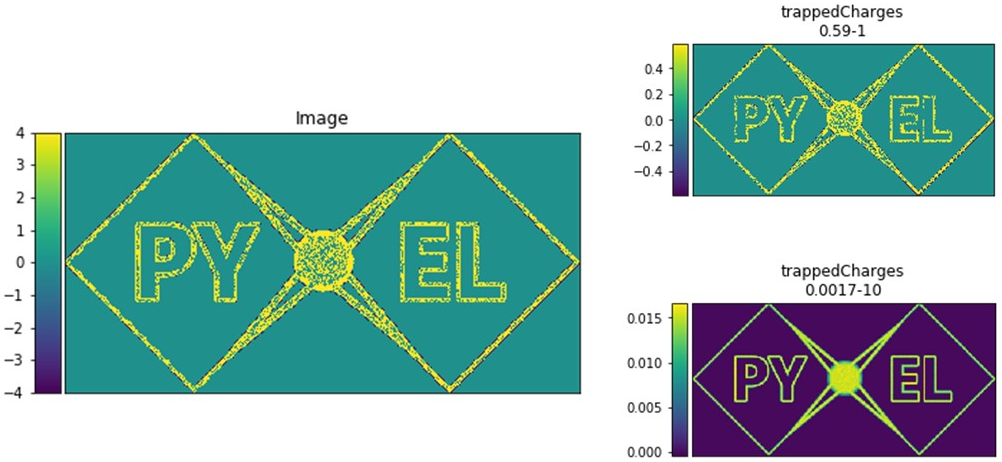}
    \caption{The detector object Image attribute (left) and memory (right) after a dark exposure following the illumination. The source was an image filled with zero to simulate a complete dark. The right top image is the trapped charge for the 1~s time constant, the right bottom one for the 10~s time constant.}
    \label{fig:persistence2}
\end{center} 
\end{figure}

We then simulate the persistence discharge by setting the detector to observe no source (complete dark conditions). The same steps as the previous simulation apply, but since the detector sees no light, the trapped charges in the memory attribute begins to decay following the persistence model and are released into the pixel for readout. This is shown in Fig.~\ref{fig:persistence2}; in the final image (left), one can see the Pyxel logo even though no light source was used in the simulation. Looking at the first memory map, we see that there is a steep decrease in the number of trapped charges. This shows that the simple persistence model that we have implemented works, as charges are stored and release from the memory like persistence would behave on a real detector.

This implementation reproduces the behaviour of persistence by using a slightly modified version of the detector object (the memory attribute of the detector object, functionality now available in version 0.7). Now that this memory attribute exists, it will be easy to implement different persistence models for testing, as the equations for charge trapping and decay would just need to be changed in the current implementation. Ongoing work on persistence at ESO has shown that a model of persistence involving the detrapping energy levels of different trap species and the operational temperatures of the detector is required to fully describe the persistence behaviour seen\cite{Ives20}; this will be the next step of the persistence model implemented in Pyxel. A working example is illustrated in one Jupyter notebook using the dynamic mode and is available in the Gitlab repository.


\section{Towards version 1.0}

In this contribution we described the motivations behind Pyxel, we summarised its working principles and gave simple how-to-install and how-to-use instructions as well as examples of more advanced usage: such as MKID-array simulation, and persistence in HxRG-like detectors.

Pyxel has now reached v0.7. It is still in beta but it is stable, contains many features, and it is becoming ever easier to use and contribute to. This is reflected by the growing community and many requests for support and new model implementations.  The next significant milestone for Pyxel is version 1.0 and for this we have established a list of objectives: the license process shall be complete, Pyxel shall be available under a well-known python software repository, each pipeline step should contain at least one implemented model for each detector type, and tutorials to demonstrate the use of each running mode shall be available. We have also recently launched a user-survey that we hope will help us to better target the needs of the Pyxel community and guide future developments. 

\bibliography{report} 
\bibliographystyle{spiebib} 
\end{document}